\documentclass[preprint]{elsarticle}
\usepackage{graphicx,url,hyperref,microtype,subcaption,graphbox,amsmath}
\usepackage{latexsym,courier,upgreek,xcolor,rotating,algorithmic,commath,ulem}
\usepackage[onehalfspacing]{setspace}
\hypersetup{colorlinks,citecolor=blue,linkcolor=red,urlcolor=green}
\usepackage[displaymath,mathlines]{lineno}
\usepackage[margin=2cm]{geometry}
\modulolinenumbers[5]
\biboptions{numbers,sort&compress}

\begin{document}

\begin{frontmatter}
  
  \title{Role of pore-size distribution on effective rheology of
    two-phase flow in porous media}
  
  \author[add1]{Subhadeep Roy}
  \ead{subhadeeproy03@gmail.com}
  
  \author[add2,add3]{Santanu Sinha}
  \ead{santanu.sinha@ntnu.no}
  
  \author[add2]{Alex Hansen}
  \ead{alex.hansen@ntnu.no}
  
  \address[add1]{PoreLab, Department of Physics, University of Oslo,
    NO-0371 OSLO, Norway}
 
  \address[add2]{PoreLab, Department of Physics, Norwegian University
    of Science and Technology, N-7491 Trondheim, Norway.}
  
  \address[add3]{Beijing Computational Science Research Center, 10
    East Xibeiwang Road, Haidian District, Beijing 100193, China.}
  
  \begin{abstract}
    The flow of immiscible fluids inside a porous medium shows
    non-linearity in the form of a power law in the rheological
    properties of the fluids under steady state flow
    conditions. However, different experimental and numerical studies
    have reported different values for the exponent related to this
    power law. Here we explore how the rheological properties of the
    two-phase flow in porous media depends on the distribution of the
    pore sizes and how it affects the power-law exponent. The
    pore-size distribution controls fluctuation in the pore radii and
    their density in a porous material. We present two approaches,
    analytical calculations using a capillary bundle model and
    numerical simulations using dynamic pore-network modeling. We
    observe crossover from a non-linear to linear rheology when
    increasing the flow rate where the non-linear part is highly
    affected by the pore-size distribution. We have also carried out
    the study for different saturations of the two fluids.
  \end{abstract}

  \date{\today}
  
  \begin{keyword}
    non-linear fluid flow, two-phase flow, porous media
  \end{keyword}
  
\end{frontmatter}


\makeatletter
\def\ps@pprintTitle{%
 \let\@oddhead\@empty
 \let\@evenhead\@empty
 \def\@oddfoot{}%
 \let\@evenfoot\@oddfoot}
\makeatother

\section{Introduction}
Multiphase flow is relevant to a wide variety of different
applications which deal with the flow of multiple immiscible fluids in
single capillaries to more complex porous media \cite{b72, d92}. The
rheology of such flow is guided by a series of parameters: capillary
forces at the interfaces, viscosity contrast between the fluids,
wettability and geometry of the system, which collectively make the
flow properties different compared to single phase flow. The study of
two-phase flow is generally divided in two regimes: (i) the transient
regime and (ii) the steady-state flow. In the transient regime, one
can obtain different types of flow patterns, namely capillary
fingering \cite{lz85}, viscous fingering \cite{cw85,mfj85} and stable
displacement \cite{ltz88}, and models such as diffusion limited
aggregation (DLA) \cite{ws81} and invasion percolation \cite{ww83} are
used to describe the patterns. When the steady state sets in after the
initial instabilities, the flow properties are determined by the
global parameters such as global pressure drops, flow rates,
saturation and fractional flow \cite{v18}.

In recent years, many studies on the steady-state two phase flow on
Newtonian fluids have revealed a non-trivial rheology, that is, in the
regime where capillary forces are comparable to viscous forces, the
relation between the total flow rate $Q$ in a sample and the global
pressure drop $\Delta P$ across it differs from a linear Darcy law
\cite{d56,w86}. Instead, $Q$ increases much faster with $\Delta P$,
obeying a power law where the power-law exponent $\beta$ is larger
than $1$ \cite{tkrlmtf09, tlkrfm09, rcs11, sbdk17}.  Furthermore,
studies have also shown that it undergoes crossovers to linear regimes
at both sides of the non-linear regime, that is, at flow rates below a
threshold and at flow rates higher than another larger
threshold. Experiments by Tallakstad et al.\ for a two-dimensional
(2D) Hele-Shaw cell filled with glass beads measured this exponent
$\beta$ as $1.85$ ($\approx 1/0.54$) \footnote{The values in the
  brackets are the exponents reported in the literature, the
  reciprocals of them should be compared here, as we express our
  results as $Q$ as a power law in $\Delta P$, whereas the cited
  articles expressed $\Delta P$ as a power law in $Q$, or rather the
  corresponding capillary number Ca.}  \cite{tkrlmtf09,tlkrfm09}. For
a three-dimensional (3D) porous media, this exponent was observed to
vary between $2.2$ ($\approx 1/0.45$) and $3.3$ ($\approx 1/0.3$)
\cite{rcs11} depending on the saturation. It was later found to
converge to a certain value $\approx 2.17$ when a global yield
pressure is considered in the system, under which there is no flow
\cite{sbdk17}. By using pore-network modeling with 2D and 3D pore
networks, Sinha et al.\ found the exponent to be close to $2$ in the
non-linear regime \cite{sbdk17, sh12}. They also have reported a
crossover to linear Darcy type regime at high capillary number when
capillary forces are insignificant. Yiotis et al.\ performed
Lattice-Boltzmann simulations with stochastically reconstructed porous
system and studied the dynamics of fluid blobs in the presence of
gravity \cite{yts13}. In the steady state, they found a non-linear
regime with quadratic dependence with an exponent $2$, which is
bounded by two linear regimes at both the high and low capillary
numbers. The blobs were then studied experimentally and the non-linear
exponent was found as $1.54$ ($\approx 1/0.65$) \cite{cst15}. Very
recently, Gao et al.\ performed experiments of two-phase flow in
sandstone samples. They used x-ray micro-tomography measurements and
for a fractional flow of $0.5$ they found the exponent in the
non-linear regime to be equal to $1.67$ ($\approx 1/0.6$)
\cite{glb20}. They also reported a regime with linear Darcy type
behavior at lower capillary numbers where the conductance does not
change significantly. Further experiments by Zhang et
al.\ \cite{zbg21} explore the dependence of the exponent on fractional
flow and reported values in the range of $1.35$ ($\approx 1/0.74$) to
$2.27$ ($\approx 1/0.44$). They presented a theory that can predict
the boundary between the linear regime and the non-linear intermittent
flow regime.

A simple explanation for the observed power law relation between flow
rate and pressure drop may be found by following the arguments of Roux
and Herrmann \cite{rh87}, concerning the conductivity of a disordered
network of resistors, where each resistor has a threshold voltage to
start conducting current.  If we compare the voltage across a resistor
in this system to the pressure drop over a link in a network of pores,
and the threshold voltage to the pressure drop necessary to overcome
capillary forces due to the presence of fluid interfaces
\cite{shbk13}, we may translate the Roux and Herrmann arguments into a
language appropriate for porous media.  When the pressure drop $\Delta
P$ across a network of pores containing fluid interfaces is increased
by an amount $\dif \Delta P$, an additional number of pores ($\dif N$)
will start contributing to the flow.  This leads to an increase in the
effective conductivity of the network as more links are participating
in the flow. If correlations between the opened links are ignored, the
increase in conductivity will be proportional to the increase in the
number of opened links, $\dif K \propto \dif N$. We integrate to find
\begin{equation}
  \label{AH1}
  K(\Delta P') \propto \int_{P_c}^{\Delta P'} \dif \Delta P''=\Delta P'-P_c\;,
\end{equation}
where the lower integration limit is determined by the threshold
pressure necessary to overcome to induce flow across the network. The
flow rate is then given by
\begin{equation}
  \label{AH2}
  Q = \int_{P_c}^{\Delta P} K(\Delta P')\dif\Delta P'\propto (\Delta P-P_c)^2\;.
\end{equation}

Tallakstad et al.\ \cite{tkrlmtf09,tlkrfm09} provided another
explanation by considering the scaling of clusters that are trapped by
capillary forces. They assumed that the flow occurs in channels in
between trapped clusters. In a two-dimensional system of length $L$
under a pressure drop $\Delta P$, a cluster will be trapped if the
capillary force $p_c > \lambda_\parallel|\Delta P|/L$, where the
$\lambda_\parallel$ is the length of such a cluster. The maximum
length of a such a trapped cluster is therefore given by
$\lambda_\parallel^m = Lp_c/|\Delta P|$. By assuming the distance
between the flow channels equal to the typical cluster length, the
total number of flow channels will be $n_c =
L/\lambda_\parallel^m$. The total flow through all the channels is
therefore the number of channels multiplied by the flow rate in each
channel, which leads to $Q \propto n_c |\Delta P| \propto |\Delta
P|^2$. However, if this formalism is extended to three dimensions, it
leads to a cubic relationship, $Q \propto |\Delta P|^3$, which is in
contrary to what is observed in experiments and simulations.

Sinha and Hansen \cite{sh12} developed a mean-field theory for a
disordered network. By analytically calculating the average
rheological behavior for such a pore \cite{shbk13} and using
Kirkpatrick's self-consistent expression for the equivalent
conductivity for a homogeneous network \cite{k73}, they derived the
relationship,
\begin{equation}
  \label{eqPQ}
  Q \propto (\Delta P - P_c)^2\;.
\end{equation}

Note that the above theoretical approaches find the exponent in the
non-linear regime $\beta$ to be equal to $2$, thus hinting at
universality.

Recently, Roy et al.\ have studied the effect of the threshold
distribution on the effective rheology of two Newtonian fluids in a
capillary bundle model \cite{rhs19}. The model consists of a bundle of
parallel capillary tubes with variable diameters along their lengths
which introduce thresholds for each tube \cite{s53,s74}. For power-law
type distributions of the thresholds they find analytically and
numerically that the non-linear exponent $\beta$ can be related to
$\alpha$, the exponent for the power-law distribution, by the
relationships $\beta = \alpha + 1$ or $\beta = \alpha + 1/2$ depending
on whether the distribution starts from zero or has lower cut off
respectively. This means, for $\alpha = 1$, the uniform threshold
distribution, $\beta$ will be equal to $2$ and $3/2$ respectively for
the two cases. This study clearly hints at the non-linear exponent
depends on the distributions related to the system properties.  We
note that the capillary fiber bundle model in the form studied by Roy
et al.\ only considered variations in the flow thresholds and not
directly on the variations in the pore sizes, nor fluctuations in the
saturation.

In this article, we present a detailed study how the distribution of
pore sizes controls the effective rheology of the two-phase flow in
the steady state. We will first study analytically the capillary
bundle model, as it is an analytically tractable model for two-phase
flow and provides deeper understanding of the underlying physical
mechanism. There we will show that there exists a transition point as
the applied pressure is increased below which the relation between
flow rate and pressure drop is non-linear. We will investigate how the
degree of such non-linearity depends on the shape and the width of the
distribution. Above the transition point we observe Darcy-like linear
flow where the distribution of pore sizes do not have any affect on
the flow equation. We will then move to numerical simulation with
dynamic pore-network modeling, where the similar transition is
observed. There the variation in exponent in the non-linear regime and
the transition point is studied by varying three parameters: the
saturation of the wetting fluid, and the span and shape related to the
pore-size distribution. Finally, with a two-dimensional plane of the
non-linear exponent vs the transition point, we show how the above
three parameters control the effective rheology of two-phase flow.

\section{Capillary Fiber Bundle Model}
In this section, we will study the analytically solvable {\it
  capillary fiber bundle model\/} (CFBM) \cite{s53,s74}, which is a
prototype for a one-dimensional porous medium. The present authors
have recently explored \cite{rhs19} this model in order to explore a
transition from non-linear to a Darcy-like behavior when the pressure
gradient across the system is continuously increased. CFBM is a
hydrodynamic analog of the fiber bundle model \cite{hhp15}, which is a
disordered system driven by threshold activated dynamics and often
used as a model system to study mechanical failure under stress.

The model consists of a bundle of $N$ independent parallel tubes of
length $L$, carrying train of bubbles with different distributions of
wetting and non-wetting fluids. A global pressure drop $\Delta P$ is
applied across the bundle, creating a global flow rate $Q$ . In the
steady state, $Q$ is the sum of all the time averaged flow rates
$\langle q \rangle$ in each individual tube. The diameter of each tube
varies along the length of the tubes which makes the interfacial
forces vary as the bubble train moves along the tubes. We assume no
film flow so the fluids do not pass each other. The total length of
the sections along the tube containing the more wetting fluid (called
the wetting fluid) is $L_w$ and the total length of the sections
containing the less wetting fluid (called the non-wetting fluid) is
$L_n$. The corresponding volumes are therefore given by $\pi r^2L_w$
and $\pi r^2L_n$, where $r$ is the average radius of the capillary
tube. The saturations in each tube are then $S_w=L_w/L$ and
$S_n=L_n/L$. Each tube making up the bundle contains the same amount
of each fluid but with its own division of the fluids into bubbles.

The total volumetric flow rate $q$ in a capillary tube at any
instant of time is given by,
\begin{equation}
  \label{eq4_CFBM}
  \displaystyle
  q = -\frac{\pi r^4}{8\mu_{av}L} \Theta(|\Delta P|-p_c) (|\Delta P|-p_c)
\end{equation}
where $|\Delta P|$ is the pressure drop across the capillary tube,
$p_c$ is the instantaneous capillary pressure given by the sum of all
the capillary forces along the capillary tube due to the interfaces
and $\mu_{av}$ is the effective viscosity given by $\mu_{av} =
S_w\mu_w + S_n\mu_n$. Here $\Theta(|\Delta P|-p_c)$ is the Heaviside
function which is $0$ for negative arguments and $1$ for positive
arguments. When the pressure difference across the tube is kept fixed,
the average volumetric flow rate $\langle q\rangle$ in the steady
state can be obtained by averaging Equation \ref{eq4_CFBM} over a time
interval,
\begin{equation}
  \label{eq6_CFBM}
  \displaystyle
  \langle q \rangle = -\frac{\pi r^4}{8\mu_{av}L} {\rm sgn}(\Delta P) 
  \Theta(|\Delta P|-\gamma) \sqrt{|\Delta P|^2-\gamma^2}\;,
\end{equation}
where $\rm sgn(\Delta P)$ is the sign of the argument. If the tube is
assumed to have a sinusoidal variation in radius with amplitude $a$
about an average radius $r$ and a period of length $l$, then $\gamma$
is given by,
\begin{equation}
  \label{eq_AH_10}
  \gamma = \sqrt{\Gamma_s^2+\Gamma_c^2}\;,
\end{equation}
where
\begin{equation}
  \label{eq_AH_11}
  \Gamma_s = \sum_{j=-K}^{+K} \frac{4\sigma a}{r}\sin\left(\frac{\pi \Delta x_j}{l}\right)\sin\left(\frac{2\pi(x_j-x_0)}{l}\right)\;,
\end{equation}
and
\begin{equation}
\label{eq_AH_12}
\Gamma_c=\sum_{j=-K}^{+K} \frac{4\sigma a}{r}\sin\left(\frac{\pi \Delta x_j}{l}\right)\cos\left(\frac{2\pi(x_j-x_0)}{l}\right)\;.
\end{equation}
Here $x_j$ is the position of center of the $j$th bubble if the tube
is filled with $2K+1$ bubbles.  Its width is $\Delta x_j$.  The
surface tension times the average contact angle is $\sigma$.

We are interested in the effect due to the variation in the pore radii
assuming that the average contribution due to the bubble sizes are the
same for all tubes. In such a scenario, the $\gamma$ can be expressed
in terms of the link radius $r$ as,
\begin{equation}
  \label{eq1_CFBM}
  \displaystyle
  \gamma = \frac{k}{r}
\end{equation}
where $k$ is a proportionality constant. Equation \ref{eq1_CFBM}
implies that the higher the radius of a tube, the lower the threshold
pressure will be and the fluids will start flowing at a relatively
lower value for $|\Delta P|$.

We now consider a bundle of $N$ fibers with average radii drawn from a
distribution $\rho(r)$.  We assume there to be a smallest radius
$r_{\min}$ and a largest radius $r_{\max}$, so that $\rho(r)=0$ for
$r<r_{\min}$ and for $r>r_{\max}$.  This means that there is a
smallest threshold $P_m=k/r_{\max}$ and a largest threshold
$P_M=k/r_{\min}$ among the $N$ fibers as $N\to\infty$.

Let us define a radius
\begin{equation}
\label{eq_AH_13}
r_c(|\Delta P|)=\max\left(\frac{k}{|\Delta P|},r_{\min}\right)\;.
\end{equation}
The total flow rate is then given by
\begin{equation}
\label{eq3_CFBM}
\frac{Q}{N}=\left\{ \begin{array}{ll}
     0 & \mbox{if $|\Delta P| < P_m$}\;,\\
     \int_{r_c(|\Delta P|)}^{r_{\max}} q\ \rho(r)\dif r & \mbox{if $|\Delta P| \ge P_m$}\;,\\
                     \end{array}
   \right.
\end{equation}
which combined with Equations \ref{eq1_CFBM}, \ref{eq6_CFBM} and \ref{eq3_CFBM} give
\begin{equation}
  \label{eq7_CFBM}
  \frac{8\mu_{av}LQ}{N\pi} = \left\{ \begin{array}{ll}
    0 & \mbox{if $|\Delta P| < P_m$}\;,\\
    -\int_{r_c(|\Delta P|)}^{r_{\max}} r^4 \sqrt{|\Delta P|^2-\left(\frac{k}{r}\right)^2} \rho(r) \dif r & \mbox{if $|\Delta P| \ge P_m$}\;.\\
  \end{array}
  \right.
\end{equation}

\subsection{Uniform distribution}
We use a uniform distribution of $r$ between $r_{\min}>0$ and
$r_{\max}>r_{\min}$ as a first illustration.  We have that
\begin{equation}
  \label{eq8_CFBM}
  \rho(r) = \left\{\begin{array}{ll}
                                0       & \mbox{if $r \le r_{\min}$}\;,\\
                                1/(r_{\max}-r_{\min})     & \mbox{if $r_{\min} < r \le r_{\max}$}\;,\\
                                0       & \mbox{if $r > r_{\max}$}\;.\\
                   \end{array}
            \right.
\end{equation}

Equation \ref{eq7_CFBM} then gives for $|\Delta P|>k/r_{\max}$
\begin{equation}
  \label{eq15-a_CFBM}
  Q=-\frac{k^5N\pi}{120\mu_{av}[r_{\max}-r_c(|\Delta P|)]L}\ \frac{1}{|\Delta P|^4} \left[(u^2-1)^{3/2}(2+3u^2)\right]_{r_c(|\Delta P|)|\Delta P|/k}^{r_{\max}|\Delta P|/k}\;.
\end{equation}
We now have two possibilities, either $|\Delta P| > k/r_{\min}$,
making $r_c(|\Delta P|)|\Delta P|/k=r_{\min}|\Delta P|/k$.  The other
possibility is that $|\Delta P| < k/r_{\min}$ making $r_c(|\Delta
P|)|\Delta P|/k=1$.

We consider the $|\Delta P| > k/r_{\min}$ case first. This is when
there is flow in all the fibers. We get
\begin{equation}
  \label{eq15_CFBM}
  Q = -\frac{k^5N\pi}{120\mu_{av}[r_{\max}-r_{\min}]L}\ \frac{1}{|\Delta P|^4} \left[(u^2-1)^{3/2}(2+3u^2)\right]_{r_{\min}|\Delta P|/k}^{r_{\max}|\Delta P|/k}\;.
\end{equation}
For large pressure drops $|\Delta P| \gg k/r_{\min}$, this expression
reduces to
\begin{equation}
  \label{eq15-b_CFBM}
  Q=-\frac{N\pi[r_{\max}^5-r_{\min}^5]}{40\mu_{av}[r_{\max}-r_{\min}]L}\ |\Delta P|\;.
\end{equation}

We now consider the opposite case, i.e.\ when $|\Delta P| <
k/r_{\min}$.  In this case, $r_c(|\Delta P|)|\Delta
P|/k=r_{\min}|\Delta P|/k$. The flow rate is then
\begin{equation}
  \label{eq15-c_CFBM}
  Q=-\frac{k^5N\pi \left[\left(\frac{r_{\max}|\Delta P|}{k}\right)^2-1\right]^{3/2}\left[2+3\left(\frac{r_{\max}|\Delta P|}{k}\right)^2\right]}
  {120\mu_{av}[r_{\max}-k/|\Delta P|]L |\Delta P|^4}\ 
\;.
\end{equation}

If we now assume that $|\Delta P|-k/r_{\max}=|\Delta P|-P_m\ll P_m$,
we may expand this expression in terms of $|\Delta P|-P_m$, finding to
lowest order
\begin{equation}
  \label{eq15-d}
  Q=-\frac{\sqrt{2}\ r_{\max}^{11/2} N\pi}{12\mu_{av}(r_{\max}-r_{\min})k^{1/2}L}\ (|\Delta P|-P_m)^{3/2}\;.
\end{equation}

\subsection{Power law distribution}
We now consider a power law distribution where link radii are chosen
with different probabilities depending on the slope of the
distribution. The expression for $\rho(r)$ we assume to be
\begin{equation}
  \label{eq17_CFBM}
  \rho(r) = \left\{\begin{array}{ll}
                                0       & \mbox{if $r \le r_{\min}$}\;,\\
                                \frac{1-\alpha}{r_{\max}^{1-\alpha}-r_{\min}^{1-\alpha}}\ r^{-\alpha}  & \mbox{if $r_{\min} < r \le r_{\max}$}\;,\\
                                0       & \mbox{if $r > r_{\max}$}\;.\\
                   \end{array}
            \right.
\end{equation}
The uniform distribution (Equation \ref{eq8_CFBM}) illustrated in the
previous section is a special case of this distribution with
$\alpha=0$. The global flow rate obtained from Equation \ref{eq8_CFBM}
is
\begin{equation}
  \label{eq18_CFBM}
  \displaystyle
  Q = -\frac{k(1-\alpha)N\pi}{8\mu_{av}[r_{\max}^{1-\alpha}-r_{\min}^{1-\alpha}]L} 
\int_{r_c(|\Delta P|)}^{r_{\max}} r^{4-\alpha} \sqrt{\left(\frac{|\Delta P|}{k}\right)^2-\left(\frac{1}{r}\right)^2} \dif r\;.
\end{equation}

As for the uniform distribution, we have two cases to consider:
$|\Delta P| > k/r_{\min}$ for which $r_c(|\Delta P|)=r_{\min}$ and
$|\Delta P| < k/r_{\min}$ for which $r_c(|\Delta P|)=k/|\Delta P|$.

We consider the case $|\Delta P| > k/r_{\min}$ first. Then, there is
flow in all the fibers and we have
\begin{equation}
  \label{eq18-a_CFBM}
  \displaystyle
  Q = -\frac{k^{5-\alpha}(1-\alpha)N\pi}{8\mu_{av}[r_{\max}^{1-\alpha}-r_{\min}^{1-\alpha}]L|\Delta P|^{4-\alpha}} 
\int_{r_{\min}|\Delta P|/k}^{r_{\max}|\Delta P|/k} u^{3-\alpha} \sqrt{u^2-1} \dif u\;.
\end{equation}
When the pressure drop $|\Delta P|$ becomes very large, i.e., $|\Delta
P| \gg k/r_{\min}$, the integral in Equation \ref{eq18-a_CFBM}
simplifies by having $\sqrt{u^2-1}\to u$ in the integrand, and we find
\begin{equation}
  \label{eq18-b_CFBM}
  \displaystyle
  Q = -\frac{(1-\alpha)[r_{\max}^{5-\alpha}-r_{\min}^{5-\alpha}]N\pi}{8(5-\alpha)\mu_{av}[r_{\max}^{1-\alpha}-r_{\min}^{1-\alpha}]L}\ |\Delta P| \;.
\end{equation}

The other case, $|\Delta P|< k/r_{\min}$ leads to $r_c(|\Delta
P|)=k/|\Delta P|$ and Equation \ref{eq18_CFBM} becomes
\begin{equation}
  \label{eq18-c_CFBM}
  \displaystyle
  Q = -\frac{k^{5-\alpha}(1-\alpha)N\pi}{8\mu_{av}[r_{\max}^{1-\alpha}-r_{\min}^{1-\alpha}]L|\Delta P|^{4-\alpha}} 
\int_{1}^{r_{\max}|\Delta P|/k} u^{3-\alpha} \sqrt{u^2-1} \dif u\;.
\end{equation}

We now assume that $|\Delta P|-P_m\ll P_m$. We expand the integral in
Equation \ref{eq18-c_CFBM} to find
\begin{equation}
  \label{eq18-e_CFBM}
  \int_{1}^{1+\Delta} u^{3-\alpha} \sqrt{u^2-1}\dif u = \int_0^\Delta (1+w)^{3-\alpha}\sqrt{2+w}\sqrt{w}\dif w = \frac{2\sqrt{2}}{3}\ \Delta^{3/2}\;,
\end{equation}
to lowest order in $\Delta$. Hence, the total flow rate is to lowest
order in $|\Delta P|-P_m$
\begin{equation}
  \label{eq18-d_CFBM}
  \displaystyle
  Q = -\frac{\sqrt{2}(1-\alpha)N\pi r_{\max}^{11/2-\alpha}}{12\sqrt{k}\mu_{av}[r_{\max}^{1-\alpha}-r_{\min}^{1-\alpha}]L} (|\Delta P|-P_m)^{3/2}\;.
\end{equation}

We see that the exponent $\alpha$ does not affect the exponent $\beta$
for $|\Delta P|$ larger than but close to $P_m$. Furthermore, for
$\alpha=0$ we retrieve Equation \ref{eq15-d}.

Here $P_m=k/r_{\max}$ plays the role of a threshold pressure $P_c$
below which there is no flow, whereas $P_M=k/r_{\min}$ is the
crossover pressure at which there is flow in all fibers. This pressure
($P_M$) signals the transition to the Darcy-type flow where $Q$ is
proportional to $|\Delta P|$.

In gist, the analytical calculations with the capillary bundle model
show that as soon as the pressure drop over the fiber bundle is large
enough for flow to start, it enters a non-linear regime where the
total flow rate $Q$ is proportional to $(|\Delta P|-P_m)^{3/2}$
irrespective of the exponent $\alpha$, defined in Equation
\ref{eq17_CFBM}. When the threshold pressure $P_M$ is crossed, the
non-linear behavior subsides and we find Darcy-type flow. Mobility is
sensitive to the details of the radius distribution for low flow rates
and insensitive at higher flow rates. We like to point out here that,
unlike the radii distribution, when a distribution of pore thresholds
is considered, a quadratic regime with $\beta=2$ can also obtained for
CFBM when the distribution has no lower cutoff \cite{rhs19}. With any
non-zero lower cut-off in the thresholds, $\beta$ is also $3/2$
there. In the present study we considered the distribution of pore
radii rather than the thresholds, and for any finite pore radii, there
is always a lower cutoff in the thresholds.

\section{Dynamic Pore Network Model (DPNM)}
Pore-network modeling is a computational technique to simulate
two-phase flow in porous media where the porous matrix is represented
by a network of pores with simplified geometries. The fluid
displacements inside the pores are governed by equations for fully
developed flow. We use a dynamic pore-network model where the menisci
positions between the fluids track the flow \cite{amhb98, sgv21}. All
the pore space in this model are assigned to links and the positions
where the different links meet are indicated by nodes. The flow rate
in such a link is given by \cite{w21},
\begin{equation}
  \label{eqWB}
  q_j = -\displaystyle\frac{g_j}{l_j\mu_j}\left[\Delta p_j - \displaystyle\sum p_c(x)\right]
\end{equation}
where $\Delta p_j$ is the local pressure drop across the $j^{\rm th}$
link. The terms $l_j$, $g_j$ and $\mu_j$ respectively define the
length, mobility and effective fluid viscosity related to that
link. If $\mu_w$ and $\mu_n$ are the wetting and non-wetting
viscosities respectively, then $\mu_j = s_{n,j}\mu_n + s_{w,j}\mu_w$,
where $s_{n,j}$ and $s_{w,j}$ are wetting and non-wetting saturations
inside that link. In this study, we consider links with circular cross
sections with radii $r_j$, for which $g_j = a_j r_j^2/8$ where $a_j =
\pi r_j^2$, the cross-sectional area \cite{l64}. The interfacial
pressure due to surface tension between the fluids is indicated by
$p_c$ which is summed over all the interfaces inside the link $j$,
taking into account the direction of the capillary forces. The links
here represent the total pore space that consists of a pore throat in
between pore bodies and the variation in the link radii along its
length is therefore modeled by a sinusoidal periodic shape. The
interfacial pressure at a meniscus inside such a pore can be expressed
by a modified Young-Laplace equation \cite{shbk13},
\begin{equation}
  \label{eqPc}
  |p_c(x_k)| = \displaystyle\frac{2\sigma\cos\theta}{r_j}\left[1-\cos\left(\displaystyle\frac{2\pi x_k}{l_j}\right)\right]
\end{equation}
where $x \in [0, l_j]$, the position of a meniscus inside the $j^{\rm
  th}$ link. Here $\gamma$ and $\theta$ are the surface tension and
the contact angle for the set of fluids and pores which are kept
constant throughout the simulations. We considered $\sigma \cos
(\theta) = 0.03 {\rm N/m}$ here. We study in-compressible fluids and
therefore at every time step $\Delta t$ we have for each node $i$ from
Kirchhoff law, $\sum q_i = 0$, where the sum is over all the links
connected to $i^{\rm th}$ link. This, together with Equations
\ref{eqWB} and \ref{eqPc} construct set of linear equations for every
node. We solve these equations by conjugate gradient method
\cite{bh88} and determine the flow rates $q_j$ in every link. All the
menisci in each link are then advanced by an amount $\Delta
x_j=q_j\Delta t$. Further technical details related to the menisci
displacements can be found in \cite{sgv21}. We consider periodic
boundary conditions that lead the system to evolve to a steady state.

\begin{figure}
  \centerline{\hfill
    \includegraphics[width=0.6\textwidth,clip]{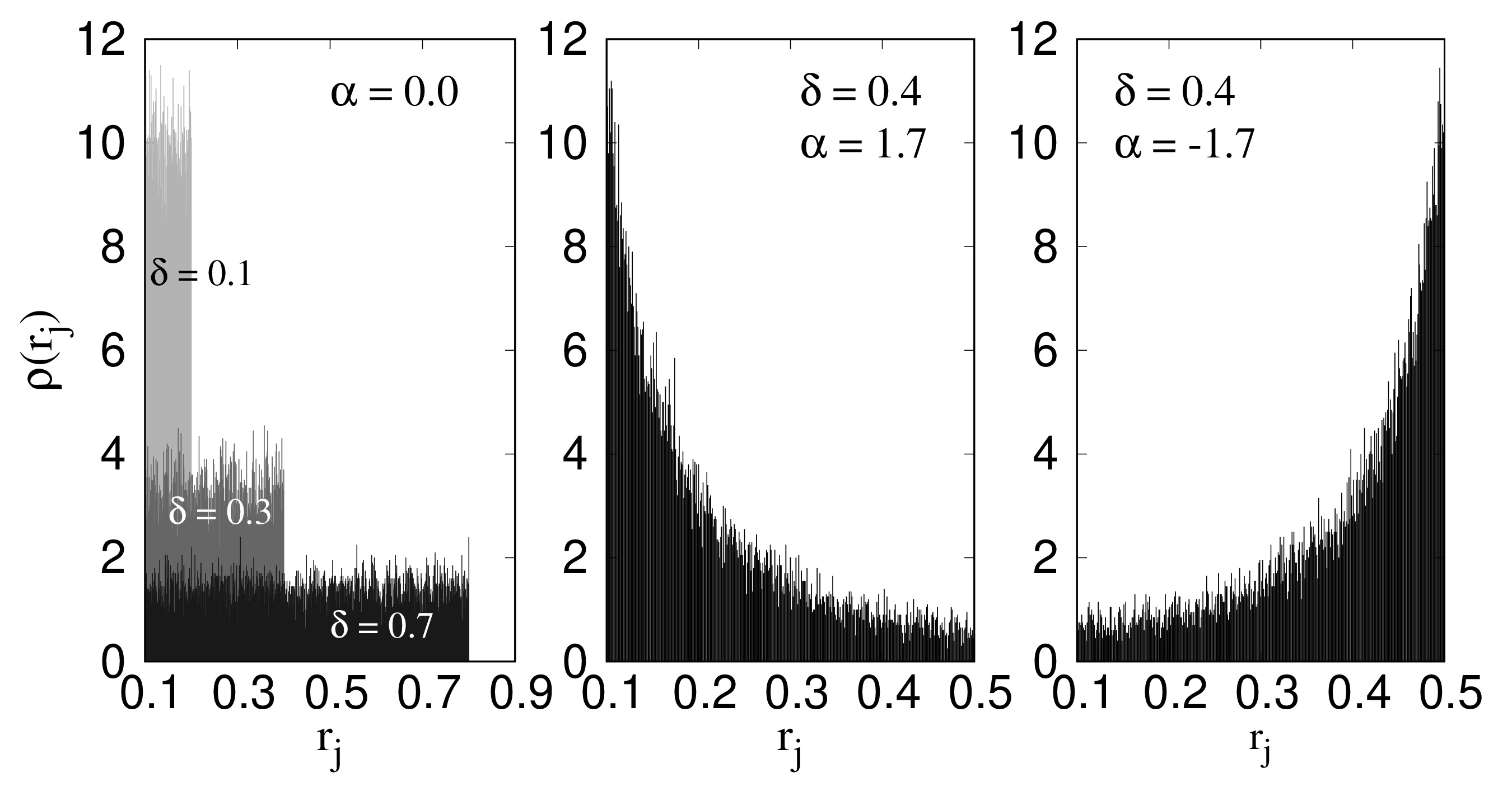}\hfill} 
  \caption{\label{figDist} Distributions of link radii $r_j$ for
    different values of the power $\alpha$ and width $\delta$. For
    $\alpha = 0$ all the radii within the interval have equal
    probability whereas for positive and negative values of $\alpha$
    there is higher probability to have narrow and wide links
    respectively.}
\end{figure}

The network we consider in this study consists of $64 \times 64$ links
in two dimensions (2D) which form a diamond lattice. All the links are
therefore at an angle $45^\circ$ with respect to the overall flow
direction. The links have equal lengths, $l_j=1{\rm mm}$, and the
disorder appears in their radii $r_j$. We choose the values of $r_j$
from a distribution $\rho(r_j)$ with a power $\alpha$ and in the range
$r_{\min}$ to $r_{\max}$ given by the same type of distribution as in
Equation \ref{eq17_CFBM},
\begin{equation}
  \label{eqDist}
  \rho(r_j) = \left\{\begin{array}{ll}
                                0                                                            & \mbox{if $r \le r_{\min}$}\;,\\
                                \frac{1-\alpha}{r_{\max}^{1-\alpha}-r_{\min}^{1-\alpha}}\ r_j^{-\alpha}  & \mbox{if $r_{\min} < r \le r_{\max}$}\;,\\
                                0                                                            & \mbox{if $r > r_{\max}$}\;.\\
                   \end{array}
              \right.
\end{equation}
We denote the width of the distribution by
$\delta=r_{\max}-r_{\min}$. The power $\alpha$ generates different
distribution types, $\alpha=0$ corresponds to a uniform distribution
whereas positive and negative values of $\alpha$ imply higher
probability to find narrower and wider links respectively. Figure
\ref{figDist} shows $\rho(r_j)$ for $\alpha=0.0$, $1.7$ and
$-1.7$. For $\alpha=0$, we show the distributions for $\delta=0.1$,
$0.3$ and $0.7$.

\section{Numerical results from DPNM}
\begin{figure}[ht]
  \centerline{\hfill
    \includegraphics[width=0.6\textwidth,clip]{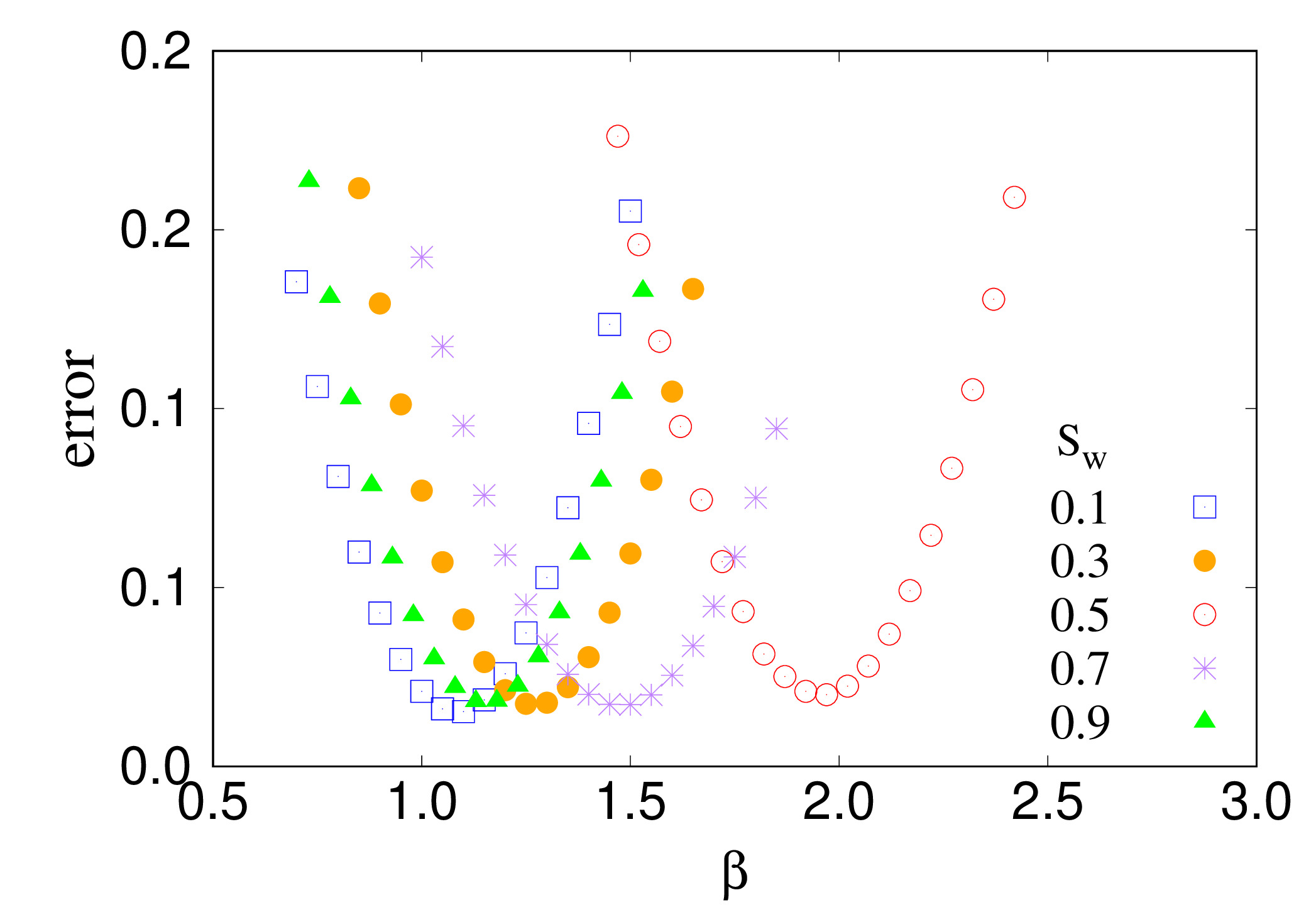}\hfill} 
  \caption{\label{figMin} The error associated with least-square
    fitting of the numerical results to Equation \ref{eqPQ} as a
    function of the trial values of $\beta$. Results are shown for
    five different saturations $S_w$. The minima of these plots decide
    the final values of $\beta$ and $P_c$.}
\end{figure}

We perform simulations with constant global pressure drop $\Delta P$
and evolve the systems to steady states, where the macroscopic
quantities fluctuate around a steady average. In the steady state, we
measure the total flow rate $Q$. Results are averaged over $20$
different samples of the pore network. By fitting the numerical data
with Equation \ref{eqPQ} for the low capillary number regime, we
calculate the exponent $\beta$ and the threshold pressure $P_c$. As
there are two parameters to determine from each data set, we
considered a method of minimizing the error related to the least
square fit \cite{sh12}. We illustrate this procedure briefly
here. First we choose a trial value of $\beta$ and perform least
square fitting with the data points. This will provide a value of
$P_c$ and an error associated with it. We perform this for a set of
$\beta$ values and find the corresponding error values. We then plot
the errors as a function of $\beta$. This is shown in figure
\ref{figMin} for five different saturations where we see non-monotonic
behavior of the errors with a minimum. We then consider the value of
$\beta$ that corresponds to the minimum error and use the
corresponding value of $P_c$. When plotting $\log(\Delta P-P_c)$ with
$\log Q$, we find the crossover point between a non-linear and a
linear regime from eye approximation. From this crossover, we then
identify the crossover pressure $P_t$.

In the following we present the results showing how the steady-state
rheology depends on the three parameters, (A) the wetting saturation
$S_w$, (B) the power $\alpha$ and (C) the width $\delta$ related to
the pore-size distribution. We keep $r_{\min}=0.1$ and vary $r_{\max}$
from $0.2$ to $0.8$, so $\delta$ varies from $0.1$ to $0.7$. The
exponent $\alpha$ is varied in the range $-1.7 \le \alpha \le 1.7$. We
will focus on the non-linear exponent $\beta$ and the pressure $P_t$
related to the cross-over from non-linear to linear regime. In a $P_t$
vs $\beta$ plane we will then highlight how the two quantities vary
when we change the values of $S_w$, $\alpha$ and $\delta$.

\subsection{Effect of $S_w$}
\label{secSw}
\begin{figure}[t]
\centerline{\includegraphics[width=0.95\textwidth,clip]{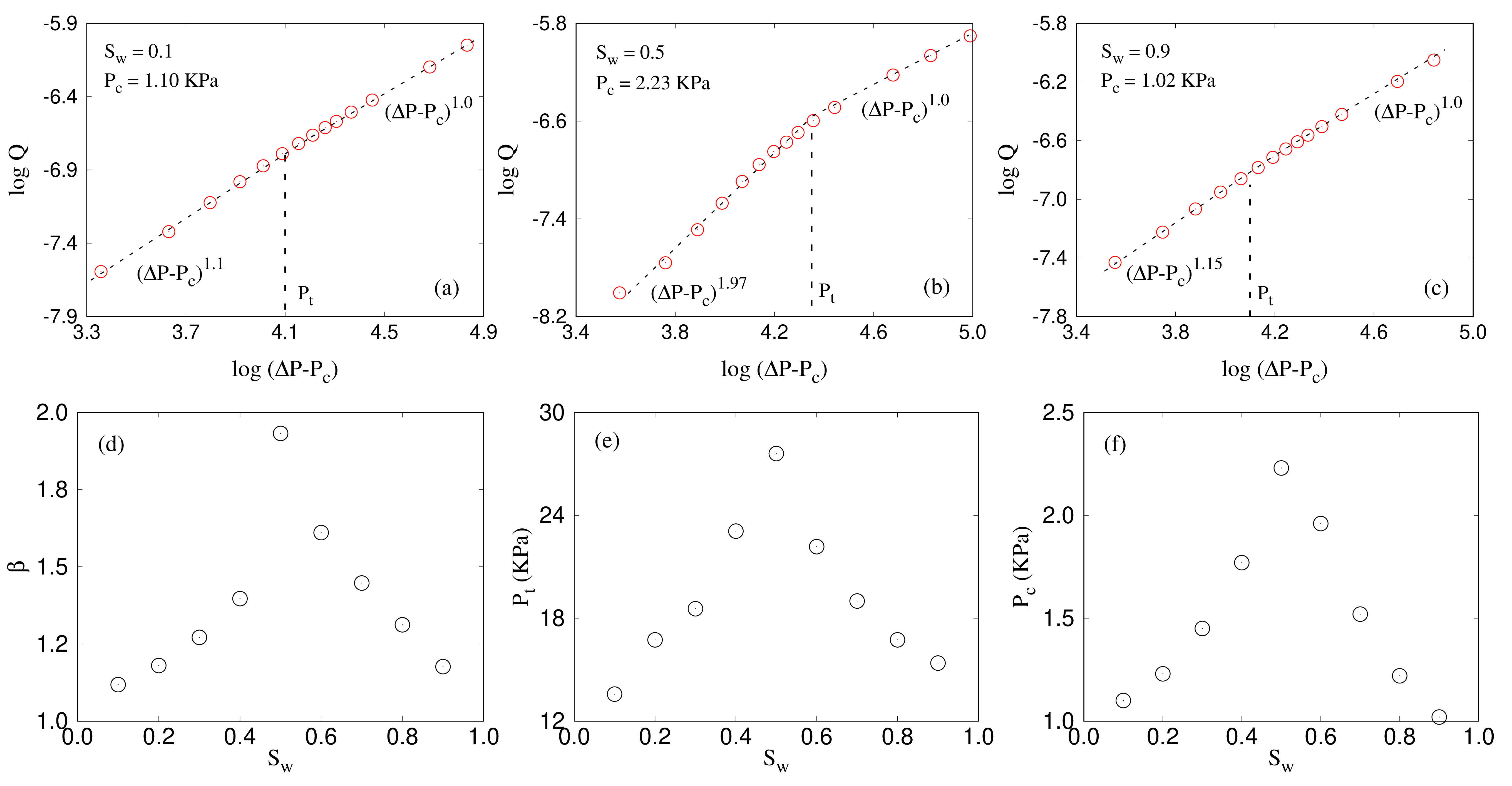}} 
\caption{\label{figSw} Plot of $\Delta P-P_c$ (Pa) vs $Q$ (${\rm
    m}^3{\rm /s}$) for a network of $64 \times 64$ links in 2D for
  three different wetting saturations $S_w = 0.1$, $0.5$ and $0.9$ are
  shown in (a), (b) and (c) respectively. The plots show a non-linear
  regime at low pressure drop and a liner regime at high pressure
  drop. The variations of the slopes $\beta$ in the non-linear regime,
  the crossover pressure $P_t$ (KPa) and the threshold pressure $P_c$
  (KPa) as a function of $S_w$ are shown in (d), (e) and (f)
  respectively. All the $\beta$, $P_t$ and $P_c$ have a maximum around
  $S_w=0.5$ and decreases on either side.}
\end{figure}

The variation of the flow rate $Q$ with the pressure drop $\Delta P$
is shown in Fig. \ref{figSw} for three different wetting saturations
(a) $0.1$, (b) $0.5$ and (c) $0.9$ where we plotted $\log(Q)$ as a
function of $\log(\Delta P-P_c)$. The value of $\alpha$ and $\delta$
are kept constant here at $0.0$ and $0.3$ respectively. The respective
threshold pressures ($P_c$), measured by the minimum error method, are
indicated in the plots. All the plots show a non-linear regime at
lower pressure drops and then a crossover to a linear regime. However,
the slope in the non-linear regime is much higher for $S_w=0.5$ than
$0.1$ or $0.9$. Similarly, the threshold pressure $P_c$ is also higher
for $S_w=0.5$. This indicates the fact that as $S_w \to 0$ or $S_w \to
1$, the two-phase flow essentially approach to the single phase flow
and it should eventually follow the linear Darcy law without any
threshold pressure. In order to see the nature of this variation
towards the linear regime, we plot in Fig. \ref{figSw} (d), (e) and
(f), the values of $\beta$ and $P_t$ and $P_c$ as a function of $S_w$,
where $S_w$ is varied from $0.1$ to $0.9$ in the interval of
$0.1$. All three plots show a peak around $S_w=0.5$ with $\beta=1.97$
and then continuously decrease in both sides. The decrease in both
$\beta$ and $P_t$ suggests that non only the non-linearity gradually
disappears as $S_w$ deviates from $0.5$, but also the linear Darcy
regime can be obtained at a relatively lower pressure drop. At the
same time, the overall threshold pressure $P_c$ also decreases as the
saturation approaches to $0$ or $1$.

If we consider that the flow occurs in channels with capillary
barriers \cite{rh87}, that is, the increase in $Q$ with $\Delta P$ is
contributed from two factors, the increase in the number of conducting
flow paths and the increase in the flow in each path, then that
explains the reduction in both $\beta$ and $P_t$ as $S_w \to 0$ or
$1$. As the saturation of a certain fluid decreases, either it reduces
the number of fluid-fluid interfaces or produces smaller bubbles of
one fluid. In both the cases the effective capillary barriers
corresponding to any flow path decreases. This results in more flow
paths with one fluid or with negligible capillary barriers which will
start flowing as soon as any pressure drop applied, making the $\beta$
to move towards $1$. At the same time, the maximum capillary barrier
that the model needs to overcome to make all possible paths flowing
also decreases, which moves the non-linear to linear transition point
to a lower value of $\Delta P$. Experimental observation of the
variation of $\beta$ with saturation was first reported in
\cite{rcs11}. No threshold pressure was considered in that study while
analyzing the results in that study. A recent experimental study
\cite{zbg21} explores the variation of $\beta$ and the crossover point
as a function of fractional flow, $F_w=Q_w/Q$. By balancing the
surface energy to create fluid meniscus to the injection energy, they
developed a theory that can predict the crossover point between the
two regimes they have studied. However, they have studied the
crossover from the linear regime at very low pressure drop to the
non-linear regime at the intermediate pressure drop, whereas our
present study addresses the crossover from the non-linear regime at
the intermediate pressure drop to the linear regime at high pressure
drop.

\begin{figure*}[t]
  \centerline{\includegraphics[width=0.95\textwidth,clip]{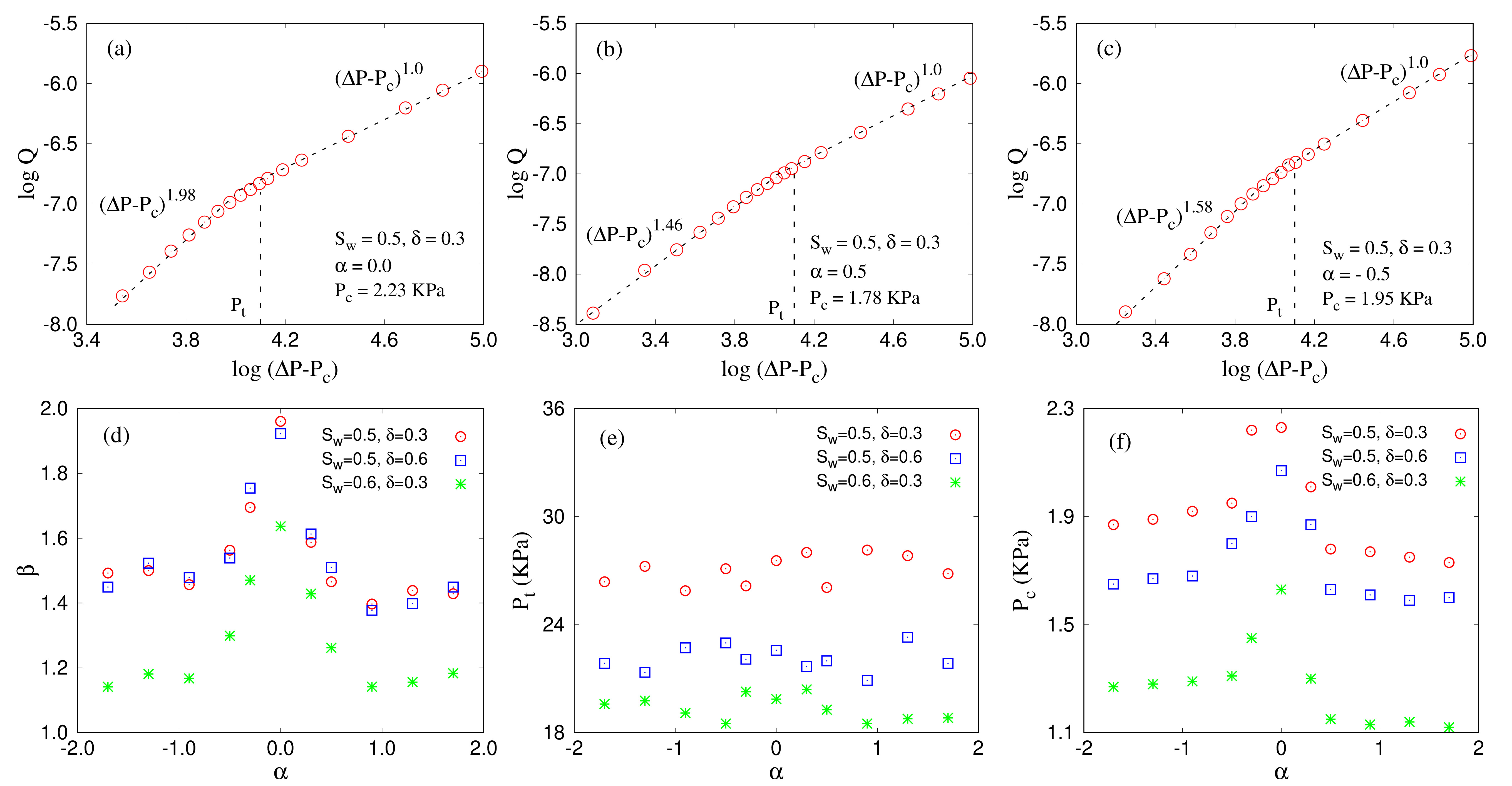}}
  \caption{\label{figAlpha} Variation of $Q$ (${\rm m}^3{\rm /s}$) as
    a function of $\Delta P-P_c$ (Pa) for a network of $64 \times 64$
    links and for $\alpha = 0$, $+0.5$ and $-0.5$ are shown in (a),
    (b) and (c) respectively. The wetting saturation $S_w=0.5$ and the
    distribution width $\delta=0.3$ here. The variations of the slopes
    $\beta$ in the non-linear regime, the crossover pressure $P_t$
    (KPa) and the threshold pressure $P_c$ (KPa) as a function of
    $S_w$ are shown in (d), (e) and (f) respectively where $\beta$ has
    a maximum at $\alpha = 0$ whereas $P_t$ does not show any specific
    dependence on $\alpha$. $P_c$ has a maximum around $\alpha=0$ and
    decreases on either side.}
\end{figure*}

\subsection{Effect of $\alpha$}
\label{seqAlpha}
The power $\alpha$ related to the pore-size distribution function
defined in Equation \ref{eqDist} determines the shape of the
distribution and tells us how the probability of having the wider
pores are compared to the narrower ones. All the pore radii within the
range $\delta$ are equally probable for $\alpha=0$ whereas positive
and negative values of $\alpha$ indicate lower or higher probability
of having wider pores respectively (Fig. \ref{figDist}). As the
interfacial pressures are inversely proportional to $r_i$ (Equation
\ref{eqPc}), the local capillary barriers are larger for $\alpha >0$
compared to $\alpha <0$. In Fig. \ref{figAlpha} we plot $\log(\Delta
P-P_c)$ with $\log(Q)$ for three different values of $\alpha$, (a)
$0$, (b) $0.5$ and (c) $-0.5$. The wetting saturation and the
distribution width are kept constant here at $S_w=0.5$ and
$\delta=0.3$ respectively. A few things are to be noticed. The
non-linear regime here is highly influenced by the value of $\alpha$
whereas the slope in linear regime remains at $\approx 1$ independent
of $\alpha$. The exponent $\beta$ is maximum for $\alpha = 0$, and
then falls to $1.46$ and $1.58$ at $\alpha = 0.5$ and $-0.5$
respectively. Moreover, the threshold pressure $P_c$ also decreases
with the increase of $|\alpha|$. The decrease in $P_c$ for $\alpha<0$
is intuitive, since the wider pores are higher here which will cause
less capillary barrier to start the flow. The decrease in $P_c$ for
$\alpha>0$ is rather counter intuitive and may be related to the
decrease in slope in the non-linear regime.

In Fig. \ref{figAlpha} (d), we plot $\beta$ as a function of $\alpha$
which shows the decreasing trends of $\beta$ on both sides of
$\alpha=0$ and then becomes constant after it reaches to $\approx 1.5$
around $|\alpha|=1$. This indicates that any change in the
fluctuations among the link radii than the uniform distribution causes
slower increase in the conductive paths when increasing the $\Delta
P$. The crossover pressure $P_t$ is plotted in Fig. \ref{figAlpha} (e)
which shows that $P_t$ remains constant around $27{\rm KPa}$
independent of the value or sign of $\alpha$. Fig. \ref{figAlpha} (f)
shows that $P_c$ shows a maximum at $S_w=0.5$ and decreases on either
side. The decrease of $P_c$ for negative $\alpha$ is understandable as
there will more links with larger radius in this case. For positive
$\alpha$ we have less links with large radius. $P_c$ is still observed
to decrease, most probably because of the nature of the fitting since
$\beta$ decreases here.

Notice that, such a variation in the non-linear exponent $\beta$ while
varying $\alpha$ was not observed in case of CFBM where $\beta$ has a
value of $3/2$ irrespective of the radii distribution (Equation
\ref{eq18-d_CFBM}). This indicates that the mixing of the fluids at
the nodes in a pore network has more complex effect than the flow in
individual channels in CFBM. The crossover point $P_t$ here is
analogous to the maximum threshold $P_M$ in CFBM which do not depend
on $\alpha$ in both the models as it only depends on the span of the
distribution and not on the the shape. Therefore, as we will see in
the next section, $P_t$ has a strong dependence on $\delta$, the span
of the distribution.

\begin{figure*}[t]
  \centerline{\includegraphics[width=1.00\textwidth,clip]{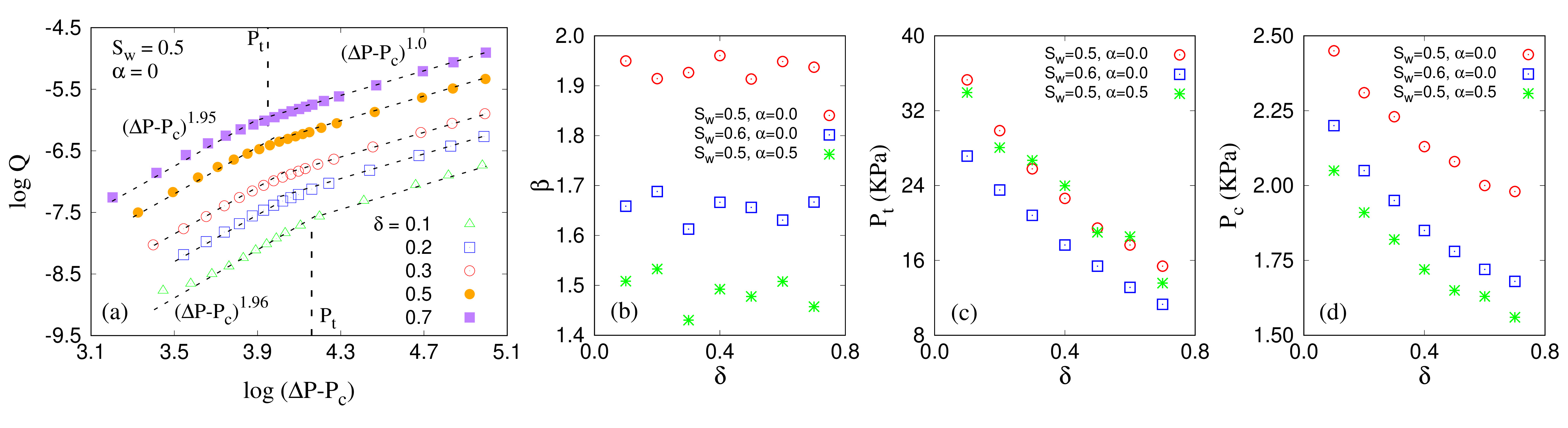}} 
  \caption{\label{figDelta} (a) Plot of $Q$ (${\rm m}^3{\rm /s}$) vs
    $\Delta P-P_c$ (Pa) when varying the distribution width $\delta$
    where the slopes in the non-linear, as well as in the linear
    regimes do not show any variation with $\delta$. In (b), (c) and
    (d) respectively, we show the variation of $\beta$ and $P_t$ (KPa)
    and $P_c$ (KPa) with $\delta$. In these results, $\alpha$ and
    $S_w$ are kept constant at $\alpha=0.0$ and $S_w=0.5$.}

\end{figure*}

\subsection{Effect of $\delta$}
\label{secdelta}
We now perform simulations by varying the width of the radii
distribution give by, $\delta = r_{\max}-r_{\min} = 0.1$, $0.2$,
$0.3$, $0.5$ and $0.7$, where $r_{\min}$ was kept constant at
$0.1$. The wetting saturation and the distribution power are kept
constant here at $S_w=0.5$ and $\alpha=0$ respectively. The results
are illustrated in Fig. \ref{figDelta}. Interestingly, unlike the
variation of $\beta$ with the distribution power $\alpha$ as seen
before, here the slopes in the non-linear regime are almost
independent of the distribution width $\delta$ and remains constant
around $2.0$ (Fig. \ref{figDelta} (a)). The crossover pressure $P_t$,
on the other hand, decreases with increase in $\delta$, which was
almost constant when we varied the distribution power $\alpha$. We
also found that the global threshold pressure $P_c$ gradually
decreases with increasing $\delta$, from $P_c \approx 2.5{\rm KPa}$ at
$\delta=0.1$ to $\approx 1.5{\rm KPa}$ at $\delta=0.7$. This is
because as we increase $\delta$, $r_{\max}$ also increases and the
system contains more pores with larger pore radii. This decreases the
capillary barriers and hence reduces the global threshold pressure
$P_c$.

\subsection{The $P_t$ vs $\beta$ plane}
\begin{figure}[ht]
  \centerline{\includegraphics[width=0.5\textwidth,clip]{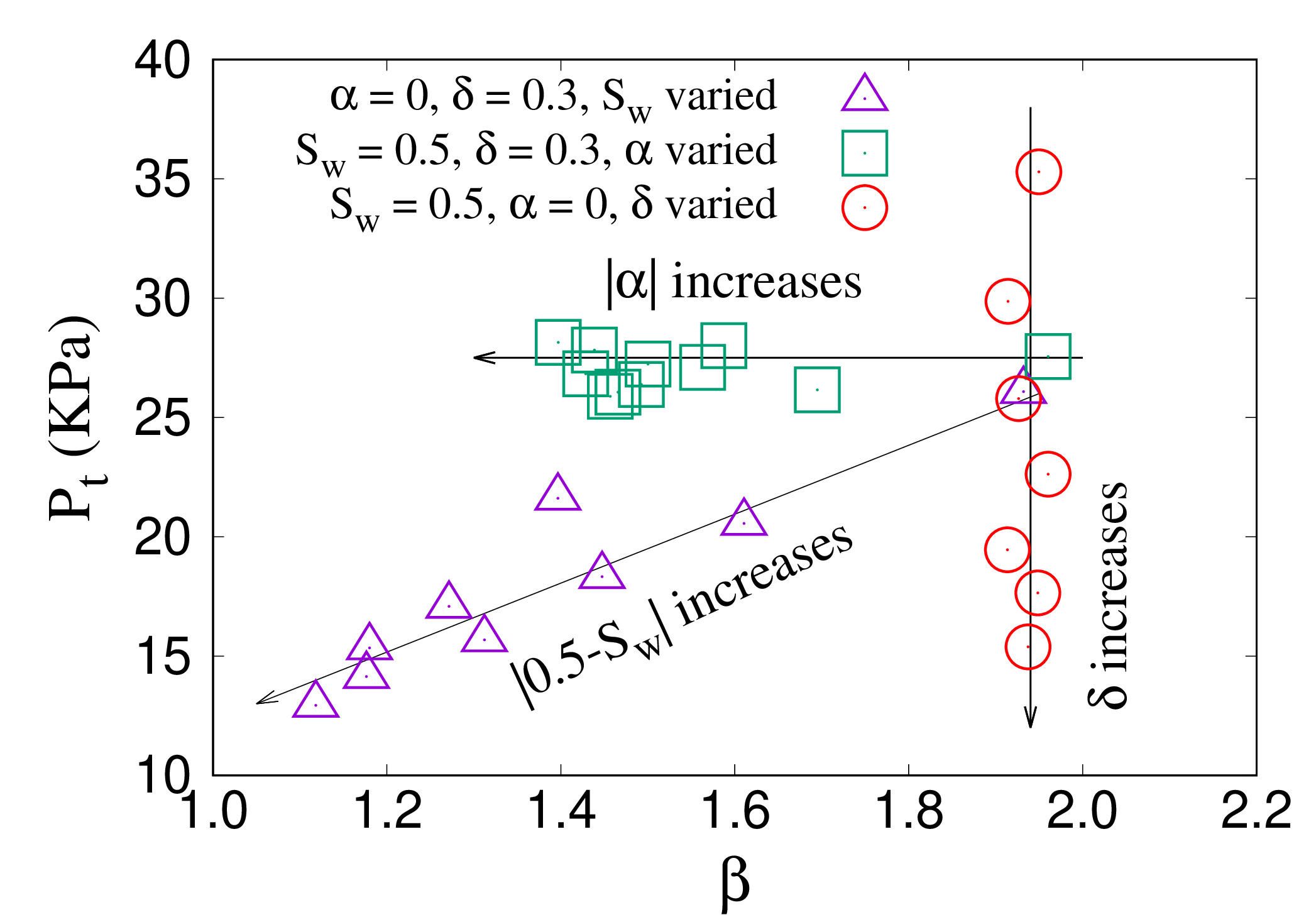}} 
  \caption{\label{figPlane} The $P_t - \beta$ plane, which shows the
    effect of varying $S_w$, $\alpha$ and $\delta$ on the rheological
    behavior. As $S_w$ deviates from $0.5$ (triangles), both $\beta$
    and $P_t$ decreases. On the other hand, with increasing $\alpha$
    (squares) and $\delta$ (circles) respectively, either $\beta$ or
    $P_t$ decreases keeping the other one constant.}
\end{figure}

To illustrate the all variations in the rheological behavior as a
function of different parameters, we constructed a $P_t - \beta$
plane. This is shown in Fig. \ref{figPlane}. We indicate all the data
points there corresponding to the three sets. The variation in the
$\beta$ and $P_t$ as we increase $S_w$, $\alpha$ and $\delta$ are
indicated by arrows in the plot. They can be summarized as:

{\sl Varying $\alpha$}: When $\alpha$ is increased, $\beta$ decreases
keeping $P_t$ constant, and the points moves to left along a
horizontal line (green squares).

{\sl Varying $\delta$}: When $\delta$ is increased, $P_t$ decreases
keeping $\beta$ constant, and the points move down along a vertical
line (red circles).

{\sl Varying $S_w$}: When $S_w$ deviates from $0.5$, both $\beta$ and
$P_t$ decreases and the points move to lower values along a diagonal
line (purple triangles).

\section{Discussion}
In this article we explored how the steady-state two-phase flow in
porous media is affected by the underlying system disorder. We
performed analytical calculations with a capillary bundle model and
numerical simulations with dynamic pore-network model. By considering
pore-radii distributions with a power $\alpha$ and a width $\delta$ we
studied the transition from a non-linear flow regime to a linear Darcy
flow. There we measured the exponent $\beta$ related to the non-linear
flow and cross-over pressure $P_t$ to the linear flow. We see that the
exponent $\beta$ is affected by $\alpha$ in the dynamic network model
but not in the capillary fiber bundle model. The crossover point is
affected by the width $\delta$ in both models. As $|\alpha|>0$ and we
deviate from a uniform radii distribution, the slope in the non-linear
regime decreases from $2$ in the dynamic network model as it decreases
the fluctuation in the pore-radii. On the other hand, when the width
of the pore size distribution $\delta$ is increased, the linear region
becomes achievable at relatively lower pressure gradient. When $S_w$
deviates from $0.5$, the two-phase flow moves closer to a single phase
flow. In such circumstances, both $\beta$ and $P_t$ decreases
simultaneously making the linear region more and more prominent. These
numerical results can be explained qualitatively with the hypothesis
that the total flow rate is contributed from the number of conducting
paths and the flow in those paths \cite{rh87}, which makes it to
increase faster than a linear behavior.

It is worth mentioning here that the single phase flow of Bingham
fluid in porous media also shows similar non-linearity with a global
yield threshold and a crossover to a linear regime \cite{rh87, ct15,
  tb13, ct15}. Nash and Rees \cite{nr17} has studied Binghum fluid
flow in 1d and obtained different relations between applied pressure
gradient and Darcy velocity depending on the distribution of channel
widths. Talon et al. \cite{tah14} has analytically explored the flow
of Binghum fluid in 1d channels with aperture variation and observed
different scaling between pressure gradient and flow rate depending on
such variation in apertures. Above studies support the fact that, when
up-scaled to a certain pore-network, both Binghum flow and two-phase
flow is affected by the pore-size distribution and hence the topology
of the of the network.

The effect of pore size distribution has been explored extensively in
case of a single phase flow where the distribution of local fluid
velocity is observed to be affected by the distribution of pore-sizes
\cite{srhwg14,wlyxc16,aqbj17,ava18,slmlm20}. In the present work we
demonstrated how flow equations are affected by the pore-size
distribution during a two-phase flow. It will be really interesting to
extend the study of velocity distribution for the two phase flow and
establish a link with the flow equations through the distribution of
pores.

\section{Acknowledgment}
This work was partly supported by the Research Council of Norway
through its Centres of Excellence funding scheme, project number
262644. SS was partially supported by the National Natural Science
Foundation of China under grant number 11750110430.

\bigskip

\end{document}